# Evolutionary Learning in the 2D Artificial Life System "Avida"

Chris Adami and C. Titus Brown
W.K. Kellogg Radiation Lab 106-38
California Institute of Technology
Pasadena, CA 91125

## Abstract

We present a new tierra-inspired artificial life system with local interactions and two-dimensional geometry, based on an update mechanism akin to that of 2D cellular automata. We find that the spatial geometry is conducive to the development of diversity and thus improves adaptive capabilities. We also demonstrate the adaptive strength of the system by breeding cells with simple computational abilities, and study the dependence of this adaptability on mutation rate and population size.

## 1 Introduction

Artificial systems such as Tom Ray's tierra have opened the possibility of studying open-ended evolution in strictly controlled circumstances, allowing experiments that were previously unthinkable as the only alternative was "wetware". The study of evolution in an information-rich artificial environment requires ever larger and faster systems, and present systems are largely restricted by such limits. Distributing tierra simulations over multiple processors is not practical on a large scale, because of the non-local interaction between members of the tierran population.

We have designed a next-generation system based on an array of cells that interact only with their nearest neighbours, and an update mechanism reminiscent of 2D cellular automata. It is designed for evolution towards complexity in an information-rich environment [3] much as the tierra system, but for purposes of universality is simpler in some respects.

Concurrently, we have retained the spirit of the tierra system, in the sense that the members of the population are strings of machine–language-like instructions running on a virtual computer much like the one designed by Ray. These strings of instructions ("genomes") can be thought of as being orthogonal to the grid that marks the physical location of the string, while the interactions between the grid points are similar to those of cellular automata. The key difference, which makes the system evolvable, is that the update rules are not fixed but rather are dependent on the genome of the cells in the immediate neighborhood. The genomes on the grid are subjected to Poisson-random mutation which allows evolutionary adaptation via implicit Darwinian selection.

The strings adapt to a landscape specified by information only: "discovery" of that information (by developing the code in the genome to trigger the bonus associated with the information) typically results in a higher replication rate for the adapted string and subsequent perpetuation of the discovery. In this manner, complexity can develop in code that starts out only with the ability to self-replicate. The task learned by the strings is entirely determined by the information they encounter, and is thus entirely at the control of the experimenter.

Suppose, for example, that we specify that adding integer numbers results in a bonus for each string that accomplishes the task. After some time, the strings *will* develop code that reads integer numbers, adds them, and then writes them to the output. Clearly, such adaptive capability can be a powerful tool, since there is no fundamental limit to the complexity achievable through use of this technique, given enough evolutionary time. We believe that this method of "stochastic information transfer", from the environment into the genome of the adapted cell via mutations, observed in tierra and this system, is central to the development of complexity in living systems.

In the next section we present a brief description of our avida system, with emphasis on the strengths of the local geometry and aspects of the update system. In section three we examine the results of the local interaction in terms of genotype age and the consequent rise in diversity. We then study the adaptive process as a function of population size and mutation rate. Finally, we offer some conclusions and discuss future applications of the system.

## 2 The avida System

In avida, the physical position of a string is determined by its coordinates in a $N \times M$ grid with the topology of a torus. As in tierra, each string is a segment of computer-code written in a simple language (similar to the Intel 80x86 assembly language) running on a configurable virtual computer[1]. The language is user-defined, but must support self-replication.

Self-replication in avida occurs when the strings copy their genome into a child string. Ideally, the strings de-termine their own size, and then allocate memory accordingly; this allocated memory is attached to the end of the genome, and the string copies its computer-code into the free space. After completing the copy, a cell-division command is issued by the string, separating the genome into two identical pieces. Once the new genome is released with the cell-division command, the oldest cell within the immediate neighborhood is replaced by a new cell containing the new genome. Thus, the birth of a cell can only affect those cells directly surrounding it. As a consequence, string-string interactions are local, and information propagates accordingly.

---
[1] For purposes of comparison, we have used an instruction set and CPU structure similar to Ray's instruction set #4.

As the strings are subject to Poisson-random mutation of their genome, the process of reproduction is often corrupted by mutation of the parent strings either before or during the copy process, leading to imperfect or incomplete copies. This is the driving force of evolutionary change and diversity in the system. Interestingly, even though the only direct source of mutations is point-permutation of the genome, many of the other recognized biological mutations emerge from the copying process; these include insertion and deletion of instructions or chunks of instructions, as well as doubling of the genome. They arise much like in the tierra system, and will be studied elsewhere. For simplicity, we do not provide for an explicit cross-over mutation mechanism, nor for multiple ploidy or sexual reproduction.

Mutation rates are defined as flux rates (mutations per site per unit time) through the available genetic space (analogous to the "soup" in tierra). While organisms interact as points on the lattice, genomes reside in a pool of genetic material. This genetic material is randomly mutated at the flux rate, whether or not it is currently in use; the results of this are discussed in section three. Finally, we have abstained from the use of flawed execution of instructions. While we acknowledge that imperfect action of proteins and enzymes do occur in nature, we have not found this to be a crucial feature in this system.

Parallel execution of the strings is simulated by assigning time-slices to each cell. After execution of its time-slice, every string is in a certain state: requesting memory, copying instructions, or placing an off-spring. After a sweep of the grid in which every cell executes its allotted time slice, the avida system updates the lattice according to the state of the cell. The interaction of cells is thus akin to $K = 1$ cellular automata[2]. The time-slice is kept small in order to insure that no cell can unduly affect its surroundings beyond control of the resolver, such as by reproducing multiple times in one time slice. However, this requirement conflicts with the need to distribute bonus time-slices as reward for the correct execution of user-specified tasks[3]. The two requirements can be reconciled via the update mechanism, which keeps the *average* time-slice constant while each individual cell is given less time (punishment) or more (reward), according to its relative bonus. In principle a cell can accumulate an unlimited amount of bonus through complex operations. When the information in the genome that triggers this bonus is propagated throughout the environment by self-reproduction of the superior organism, a newly born cell without this information will be at a severe disadvantage obtaining significantly less than the average time-slice and thus unable to compete, while already existing organisms and newly born organisms that contain this information will compete on an equal footing, receiving the average slice.

The update mechanism is also designed to allow simple distribution of avida over multinode systems. Because the only direct interaction of strings is local, the transfer of information between computational nodes is kept to a minimum; in fact, the only two mechanisms that necessarily communicate between nodes are the conflict resolution system (which handles reproduction) and the time slicing mechanism. Therefore avida can be distributed over multiple processors with a nearly linear increase in execution speed.

For further information on the precise implementation of avida see [8]. The program is available to the public[4]. For a more detailed introduction to tierra-like auto-adaptive genetic systems, we refer the reader to Refs. [1-3].

## 3  Localized Interactions and Genotype Age Distribution

In this section we would like to point out the relationship between causal spread of information and the genotype-age distribution.

In avida, as in tierra, a genotype system records the creation and extinction of species via their populations. Genotypes are exact: every member of a specific genotype has the same genome and therefore most point mutations (and consequent reproductive mutations) create new genotypes. This abundance of new genotypes, while an important part of any evolutionary system, is hard to track in a meaningful way. In addition, most (more than 90%) of mutations generate nonviable genomes. These genomes die out quickly, with minimal impact on the system.

In order to observe the system more easily, we have (as in the tierra system) divided genotypes into *threshold* and *temporary* genotypes. Threshold genotypes are those that have at any time achieved at least 10 concurrent members, while all other genotypes are considered temporary. The ages of all threshold genotypes can be plotted as a frequency distribution $N(\tau)/N$, where $N$ is the total number of genotypes, to reveal information about the diversity and survival probability of genotypes.

We have measured this distribution in tierra by collecting the ages of 184,767 genotypes from 40 runs at the same mutation rate (see Fig. 1).

---

[2]We have experimented with $K = 2$ interaction with no significant change in dynamics.

[3]see Ref. [3]

[4]To obtain avida, contact one of the authors.

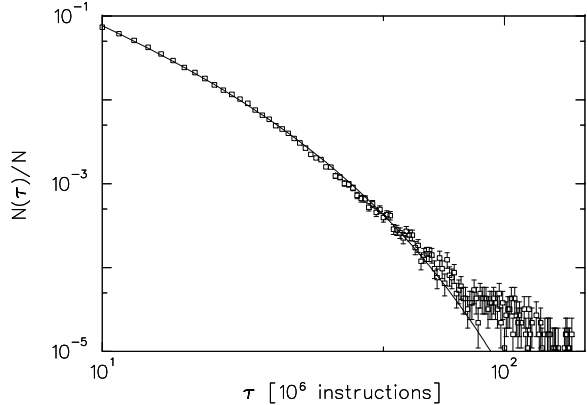

FIG. 1 Distribution of ages of genotypes in a tierran population of $\sim 1000$ cells at mutation rate $R = 0.65 \times 10^{-3}$ (mutations per instruction executed). This fit yields $D = 1.6 \pm 0.05$, with a cut-off $T = 15 \pm 1$ million instructions.

Due to the threshold in tierra, short-lived genotypes are not recorded and the distribution shows a lack of young genotypes. For that reason, we plot only genotypes that are older than 10 million instructions. We fit the distribution to a power-law $N(\tau)/N \sim \tau^{-D} \exp(-\tau/T)$ with a finite-time cut-off, and find $D = 1.6 \pm 0.05$ and $T = 15 \pm 1$ million instructions. However, the value for $D$ must be considered a lower limit as removal of points at small genotype ages (which may be affected by the threshold) decreases the $\chi^2$ of the fit and $D \to 2$.

In avida we chose a much less severe threshold and see no dependence of the fit on the removal of points at young genotype ages. The measured distribution is shown in Fig. 2.

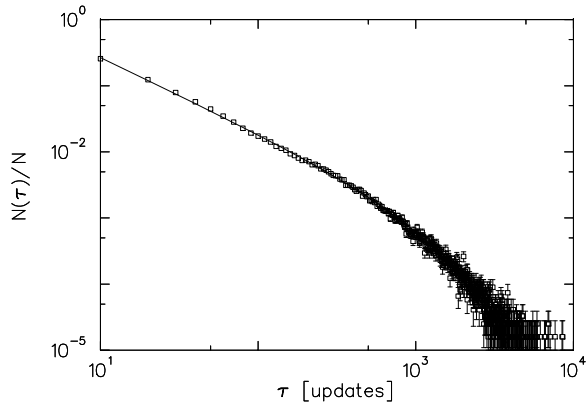

FIG. 2 Distribution of ages of genotypes in population of 1600 cells at mutation rate $R_\star = 2.0 \times 10^{-3}$. The fit yields $D = 1.14 \pm 0.1$, with a cut-off $T = 1200 \pm 20$ updates.

This is the distribution of ages for a population of 1600 strings at mutation rate $R_\star = 2 \times 10^{-3}$ mutations per executed instruction, obtained from 20 runs under identical conditions, yielding 121,703 genotypes. We fit the distribution with the same parametrization and find the slope $D$ to be $1.14 \pm 0.1$, with a cut-off $T = 1200$ updates. Note that the absolute values for the lifetimes cannot be compared easily, as the units of time are different in the two systems. The exponents, however, should be universal.

Examining the distributions further, we find $D$ to be independent of population size or mutation rate; the exponential cutoff $T$, while independent of population size, does seem to depend on the mutation rate. This is most likely due to the different average run length at different mutation rates.

The rather rapid decrease of $N(\tau)$ in tierra and the correspondingly small number of "old" genotypes suggests a tendency toward premature equilibration in tierra. Indeed, as any cell anywhere in the tierran soup can affect any other cell *directly* via the reaper queue, a discovery anywhere in the soup will reach other cells immediately and force extinction of those genotypes. Thus, diversity is throttled and the population will tend to homogenize. This is especially worrisome if the governing genotype is trapped in a meta-stable state. In this case, the lack of diversity may condemn the population to remain in this state indefinitely. This kind of behavior is apparent in tierra simulations with large soup-sizes.

The low value of $D$ in avida on the other hand suggests a near maximal population diversity. This signifies the simultaneous exploration of multiple evolutionary paths in the system, a feature of a robust Darwinian environment. It is the result of a causal spread of information, a direct consequence of the localized interaction.

Extinction events in avida are thus far less severe than extinction events in tierra and do not seriously curtail the heterogeneity of the environment. Significant gains are still disseminated throughout the population, but the new information has more time to be integrated with the existing genotypes. This strongly suggests that the meta-stable states observed in tierra [3] will not play as large a role in halting the evolution of new genotypes in the system, and indeed, very few meta-stable states have been observed.

## 4 Evolution and Adaptation

In this section we investigate the "learning" capabilities of avida as a function of external mutation rate and population size.

We design a landscape for the population to adapt to by distributing bonuses for accomplishing either the main task or certain other feats that are helpful in building up the code necessary to trigger the main bonus. Specifically, to compare with results obtained with tierra [3], we breed strings that have the ability to add two integer numbers. The code necessary to perform this task must at least include two "read" and one "write" statement, as well as register addition and movement of numbers between registers. The strings start out as self-replicating, with no other capability. With the average time-slice set to 30, we reward each "read" and "write" statement that develops with 7 time-units, for a maximum of three "read" and three "write" statements. Additional input/output instructions are not rewarded. Furthermore, if a string manages to echo the value last read into the output buffer unchanged, it reaps a bonus of 30 units for each time this task is accomplished, with a maximum of three times. Finally, if a string writes into the output buffer a number that is the sum of two previously read numbers, it is rewarded with 100 units, with a maximum repetition rate of three. Note that while highly adapted cells can reap a bonus of well over 300, the average time-slice per update remains constant.

All these bonuses are of course available at the same time, and no order is specified. Most importantly, they do not favour a particular solution to the problem but rather create a fitness gradient that leads to the solution from many avenues. Evidently, the paucity of the reward structure does constrain the solutions to a certain class while a more complex environment would allow solutions to the problem taking advantage of rewards entirely unconnected to the task at hand. This is of course a feature of evolution in natural systems, and the construction of a more complex environment is a challenging task for the future.

Since adaptation by mutation is an intrinsically stochastic process, the definition of an "average learning rate" is problematic. Intuitively, we would expect that an average learning time[5] should be connected to the average time between "discoveries" (discontinuous jumps in replication rate).

Yet, it was observed recently [4] that the time between such jumps is distributed according to a power law, and no such average can be defined. This can be traced back to the fact that there is no time scale of the order of the learning time in auto-adaptive genetic system. We can nevertheless determine the adaptive power of the population by obtaining the *learning fraction* $f_X$, which is the fraction of runs that have accomplished the task *before* a cut-off time $X$ (measured in thousands of updates). Thus, if for ten runs under identical conditions we find six where almost all of the population has discovered how to trigger the bonus before, say, 10,000 updates, this combination of parameters is assigned $f_{10} = 0.6$.

In Figs. 3a-c and Figs. 4a-c we show the learning fraction for an array of 20x20 and 40x40 cells respectively, for cut-offs $X=10$, 20, and 50 thousand updates, as a function of mutation rates. To obtain these graphs, we repeated up to 20 runs at each of 15 mutation rates between $R_\star = 0.1 \times 10^{-3}$ and $R_\star = 20 \times 10^{-3}$ [mutations per executed instruction] which translates into $R = 0.12 \times 10^{-4}$ to $R = 23.44 \times 10^{-4}$ [mutations·(site)$^{-1}$·(update)$^{-1}$]. The number of runs $N$ performed at each mutation rate is shown in Tab. 1, the total number of runs being 640.

| $R_\star[10^{-3}]$ | $R[10^{-4}]$ | $N(20\text{x}20)$ | $N(40\text{x}40)$ |
|---|---|---|---|
| 0.1 | 0.12 | - | 20 |
| 1.0 | 1.17 | 20 | 20 |
| 2.0 | 2.34 | 20 | 20 |
| 3.0 | 3.52 | 20 | 20 |
| 4.0 | 4.69 | 20 | 20 |
| 5.0 | 5.86 | 20 | 20 |
| 6.0 | 7.03 | 20 | 20 |
| 7.0 | 8.20 | 20 | 20 |
| 8.0 | 9.37 | 20 | 20 |
| 9.0 | 10.55 | 20 | 20 |
| 10.0 | 11.72 | 20 | 20 |
| 11.0 | 12.89 | 20 | 20 |
| 12.5 | 14.65 | 20 | 20 |
| 14.0 | 16.41 | - | 20 |
| 15.0 | 17.58 | - | 20 |
| 17.0 | 19.92 | 20 | 20 |
| 20.0 | 23.44 | 10 | 10 |

TAB. 1   Number of runs performed for an array of 400 (3rd column) and 1600 (4th column) cells at mutation rate $R_\star$ (in units mutations per executed instruction) [first column] or $R$ (in units mutations per site per update) [second column].

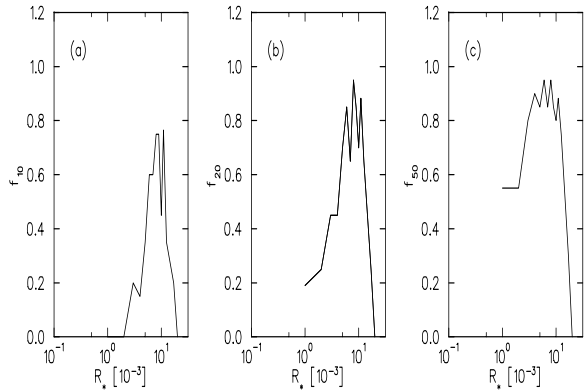

FIG. 3 Learning fraction vs. mutation rate for 400 cells in a $20 \times 20$ lattice, for three different cut-offs (a): $f_{10}$, (b): $f_{20}$, and (c): $f_{50}$.

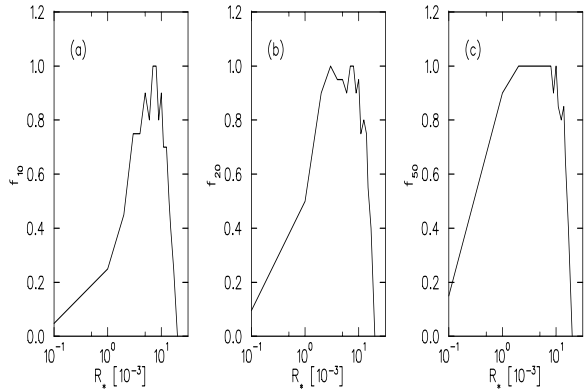

FIG. 4 Learning fraction vs. mutation rate for 1600 cells in a $40 \times 40$ lattice. Notation as in Fig. 3.

---

[5]Note that in avida, we choose the number of updates as the universal measure of time, as this is independent of population size.

While we were interested in results for larger populations as well, it turned out that runs for 6400 cells tended to be extremely CPU-time consuming if allowed to continue to the scheduled number of 50,000 updates. While most runs at the geometry 80x80 did learn the task fairly easily, we do not report any results for lack of statistics.

It is intuitively obvious that there should be an optimum mutation rate at which the strings on average achieve the task in the smallest amount of time. Clearly, very small mutation rates may take the population a long time to adapt. On the other hand, a mutation rate that is so high that the average time between mutations hitting a cell is smaller than the gestation time will prevent the information contained in the genome to be transmitted, and self-replication stops [3]. This limit is often called the "error-catastrophe" limit [6] and is crucial in understanding the window of adaptability.

Both the small and the large population simulations show the rapid drop-off as the mutation rate reaches the error-catastrophe limit. Interestingly, however, this limit is, for the genome-size we started out with (an ancestor of size 59) exceeded in most runs. This can be traced back to an evolutionary pressure to reduce genome size at high mutation rate. Indeed, we find that this pressure is intense starting at around $R_\star \approx 10^{-2}$. If the pressure to reduce genome size (and thus to represent less of a target to the lethal mutations) is high, cells tend first to reduce genome size and then learn to add. Away from the error-catastrophe limit the pressure to reduce size is less intense, but the mutation rate too high to develop the necessary code to reap the bonus. As a consequence, we witness regions in the "learning window" where learning is suppressed. Also, we see a clear effect of population size on the learning rate. The learning fraction rises earlier for larger populations, and reaches saturation (all runs learn to add integer numbers before the scheduled finish) earlier.

## 5 Conclusions

We have introduced a next-generation auto-adaptive genetic system with local interactions between members of the population and causal propagation of information on a two-dimensional flat torus (periodic boundary conditions in $x$- and $y$-directions). An update mechanism that allows the strings to be executed in any order (akin to the update of cellular automaton arrays) guarantees parallelism at all times, as the average time-slice is kept constant and small. We determined that the local interactions lead to a more diverse population with a larger spread in genotype-ages, and less liability of trapping in meta-stable states, as occurs frequently if the population is too homogenous.

We investigated the evolvability of the population as a function of population size and mutation rate, and found that the learning fraction for large arrays rises earlier and thus offers a broader window for evolution, while the upper limit (error-catastrophe limit) is universal for all sizes, as expected.

Due to the flexibility of the avida system, it can be used in many varied applications. Besides breeding strings to perform user-specified tasks, it can be used for research on evolution, ecology, and maybe immunology. As an example, we have developed a method to measure the genetic distance between genotypes [7], and are planning to use it to study trajectories in genotype-space in quasi-deterministic and chaotic regimes.

## Acknowledgements

We would like to thank Steve Koonin for continuing support, and Charles Ofria for collaboration in the design of avida. We also thank Reed College for the use of its computational chemistry laboratory. CTB acknowledges a SURF fellowship from Caltech. CA is supported in part by NSF grant PHY90-13248 and a Caltech Divisional Fellowship.